\documentclass[doublecol, subeqn]{epl2} 
\usepackage{hyperref}
\usepackage{amssymb}
\usepackage{amsmath}
\usepackage{mathrsfs}
\title{Unimodular conformal and projective relativity}
\shorttitle{Unimodular conformal and projective relativity} %Insert here a short version of the title if it exceeds 70 characters

\author{K. Bradonji\'{c}\inst{1} \and J. Stachel\inst{2}}
\shortauthor{K. Bradonji\'{c} \etal}

\institute{                    
  \inst{1} Physics Department, Boston University, Boston, MA 02215 USA\\
  \inst{2} Center for Einstein Studies, Boston University, Boston, MA 02215 USA
}

\newcommand{\ka}{\kappa}
\newcommand{\m}{\mu}
\newcommand{\n}{\nu}
\newcommand{\s}{\sigma}

\newcommand{\la}{\lambda}
\newcommand{\g}{\tens{g}}
\newcommand{\cg}{\widetilde{\tens{g}}}

\newcommand{\pp}{\overline{\nabla}}
\newcommand{\cp}{\widetilde{\nabla}}
\newcommand{\metp}{\stackrel{m}{\nabla}}
\newcommand{\ap}{\stackrel{a}{\nabla}}

\newcommand{\be}{\begin{equation}}
\newcommand{\ee}{\end{equation}}

\newcommand{\p}{\partial}

\newcommand{\f}{\frac}

\newcommand{\tdel}{\tens{\delta}}

\newcommand{\proj}{\tens{\tPi}^{\ka}_{\m\n}}
\newcommand{\ch}{\{^{\ka}_{\m\n}\}}
\newcommand{\bs}{\begin{split}}
\newcommand{\es}{\end{split}}

\newcommand{\ddd}{\hspace{2.1pt}.\hspace{2.1pt}.\hspace{2.1pt}.}

\newcommand{\R}{\mathbb{R}}

\newcommand{\dd}{\hspace{2.1pt}.\hspace{2.1pt}.\hspace{1.7pt}}

\newcommand{\SL}{SL(4,\R)}

\newcommand{\GL}{GL(4,\R)}
\newcommand{\M}{\mathcal{M}}
\newcommand{\tPi}{\tens{\Pi}}
\newcommand{\G}{\tens{\Gamma}}
\newcommand{\cpd}{\widetilde{\overline{\tens{T}}}}

\newcommand{\td}{\tens{T}}
\newcommand{\cpq}{\stackrel{\simeq}{\tens{Q}}}

\newcommand{\q}{\tens{Q}}
\newcommand{\ep}{e^{\varphi}}
\newcommand{\T}{\tens{\mathcal{T}}}
\newcommand{\Ri}{\tens{R}}

\abstract{We outline unimodular conformal and projective relativity (UCPR), an extension of unimodular relativity in which the conformal and projective structures play central roles. Under $\SL$ symmetry group, the pseudo-Riemannian metric naturally decomposes into a four-volume element and a conformal metric; and the affine connection decomposes into a one-form and a trace-free projective connection. In UCPR, these four space-time structures are treated as independent fields that have clear physical interpretations. A Palatini-type variational principle for the usual general relativity Lagrangian leads to a breakup of the Einstein field equations and the compatibility conditions between the metric and connection. We indicate how new gravitational theories may be generated by modifications of this Lagrangian and discuss two such cases. Finally, we discuss possible physical consequences of our results for quantum gravity.}

\begin{document}

\maketitle

\section{Introduction}
Unimodular relativity (UR) is a modification of general relativity (GR) in which the symmetry group of space-time is reduced from the general linear group $\GL$ to its subgroup, the special linear group $\SL$. Mathematically, this reduction has two main interrelated effects:
\begin{enumerate}
\item{It abolishes the distinction between tensors and tensor densities. For example, four-volume element scalar densities and conformal metric tensor densities become four-volume element scalars and metric tensors, respectively.}
\item{It provides an invariant breakup of certain geometric objects into their traces and trace-free parts. For example, the Levi-Civit\`{a} and affine connections break up into trace-free conformal connections plus the gradients of scalars, and trace-free projective connections plus one-forms, respectively. Similarly, the Einstein and stress-energy tensors break up into trace-free tensors and scalar traces.}

\end{enumerate}

UR has found its way into numerous approaches to classical and quantum gravity and cosmology  \cite{Einstein, Anderson, Finkelstein, Henneaux, Unruh, Weinberg, Bombelli, Daughton, Alvarez, Alvarez2, Smolin2009, Chiou, Smolin2011},  dark energy~\cite{Shaposhnikov} and even particle physics \cite{Blas}. Common to all these treatments is the breakup of the metric tensor $\g_{\m\n}$ into the \emph{conformal structure} represented by a conformal metric $\cg_{\m\n}$ with det$(\cg_{\m\n})=-1$, and a four-volume element $\m>0$ (from now on denoted by $\ep$).

But there are two major shortcomings of these approaches. First, some of them treat the four-volume element as non-dynamical. However, the restriction of the symmetry group to $\SL$ does not at all require that $e^{\varphi}$ be treated as a non-dynamical scalar field. Second, even those approaches that do dynamize the four-volume structure take the metric as the only fundamental spatio-temporal field. However, GR is based on two space-time structures that are both mathematically and physically quite distinct: a metric and an affine connection. The metric field defines the length of spatial and temporal intervals in any local inertial frame at any point of space-time; these intervals can be measured by ideal rods and clocks at rest in that frame, hence the physical characterization of the metric field as the \emph{chrono-geometry}. But it does not allow comparison of such intervals at different points of space-time. This requires an affine connection, which enables such comparisons by means of parallel transport along some path between two points. 

On its own, the affine connection determines the autoparallel paths in space-time. Physically,  these paths describe the motion of point masses ``freely falling" under the combined influence of inertia and gravitation, hence its physical characterization as the \emph{inertio-gravitational} field. The connection also defines a preferred affine parameter along these paths, which in GR may be identified with the metrical proper time. 

The specification of some chrono-geometric structure, some inertio-gravitational structure, and compatibility relation between them forms a part of any gravitational theory obeying the equivalence principle. This holds true for Newtonian and special-relativistic gravitational theories, as well as for GR. The fact that the chrono-geometry uniquely determines a torsion-free affine connection in GR does not affect this crucial conceptual distinction~\cite{EinsteinMeaning}. 

But neither the metric nor the affine connection is an irreducible structure. T. Y. Thomas first noted that the group of $\SL$ transformations simplifies the transformation laws of many geometrical objects and leads to a breakup of the affine connection $\G^{\s}_{\m\n}$~\cite{Thomas} . Namely, the affine connection breaks up into a \emph{projective structure} represented by projectively invariant, trace-free, connection $\tPi^{\ka}_{\m\n}$ that determines  paths of freely falling massive particles; and a one-form $\tens{\G}_{\m}$  associated with the preferred parameterization of these autoparallel paths.

Hence the framework of UR can be widened naturally to unimodular conformal and projective relativity (UCPR), a formulation of space-time theory in terms of four independent fields: a four-volume element $e^{\varphi}$, a conformal metric $\cg_{\m\n}$, an affine one-form  $\G_{\m}$, and a projective connection $\tPi^{\s}_{\m\n}$. This approach shows how to reach the full compatibility of the four fields in GR through a series of intermediate steps; and how these conditions, as well as the homogeneous and inhomogenous Einstein field equations, can be derived from a ``natural" decomposition of the usual Palatini-type variational principle. 

Of course, if one wants to modify GR, one need not impose full compatibility. Separate consideration of all four structures and their inter-relations allows for the possibility of introducing other Langrangians based on other $\SL$ invariants that can be formed from the four fields. 

\section{Unimodular conformal and projective relativity}

Leaving a thorough account  for a forthcoming paper~\cite{Bradonjic}, we now present a concise exposition of UCPR. We begin with a  more systematic account of the four structures defined on a four-dimensional differentiable manifold $\M$. Hereafter, all references to ``metric"  and ``affine connection" refer to a pseudo-Riemannian metric field $\g_{\m\n}$ with signature $(+, -, -, -)$, and a torsion-free affine connection field $\G^{\s}_{\m\n}$, respectively, on $\M$. Following Schouten, all symmetrization and antisymmetrization brackets carry a factor of $1/2$, and all curvature and Ricci tensors are formed from their respective connections in accord with the following conventions~\cite{Schouten1954}:
\be
\label{eq.curvature}
\tens{R}^{\ddd\ka}_{\n\m\la}\equiv2\p_{[\n}\G^{\ka}_{\m]\la}+2\G^{\ka}_{[\n|\rho|}\G^{\rho}_{\m]\la}, 
\ee
and
\be
\label{eq.ricci}
\tens{R}_{\m\la}\equiv\tens{R}^{\ddd\ka}_{\ka\m\la}.
\ee
\begin{itemize}
\item{Conformal structure and scalar field} 

In standard GR, equivalence classes of conformally equivalent metrics can be characterized by a conformally invariant metric tensor density $\cg_{\m\n}$ of weight $-1/2$  and determinant $-1$. Restricted to $\SL$, $\cg_{\m\n}$ and its dual $\cg^{\m\n}$ transform as tensors and can be used to lower and raise indices, a convention we adopt in this paper. The conformal structure defines the cone of null or ``light-like" directions at each point, and hence the null paths in space-time, as well as the dual null hypersurfaces which correspond to wave-front characteristics  (null hypersurfaces of constant phase) for any zero-rest-mass field. Classically, the wave description of the propagation of such fields is primary; the dual null ``light ray" (bicharacteristic) description is only useful in the eikonal or ``geometrical-optics" approximation~\cite{Benenti}. As we explain in the conclusion, actual interaction of field quanta (e.g., photons) with matter cannot be described in this way.

One can generate any member of the conformal equivalence class by multiplying the conformal metric by a scalar field $e^{\varphi}$, resulting in a new metric 
\be
\g_{\m\n}=e^{\varphi}\cg_{\m\n}\tx{, }\hspace{.2in}det(\tens{\g_{\m\n}})=-\g=-e^{4\varphi}.
\ee

$\SL$ can act only in a way that preserves the four-volume element, the magnitude of which depends on $\varphi$.  Hence, $\varphi$ is not a given quantity, but potentially as much a dynamical field as $\cg_{\m\n}$. We agree with the suggestion that, in the classical continuum limit, this structure may be the remnant of a more fundamental discrete quantum structure of space-time itself~\cite{Anderson, Daughton}; but in UCPR this no longer implies that it must be a non-dynamical variable. We discuss some implications of this for quantum gravity theory in the concluding section.

There are two covariant derivatives associated with the conformal structure: the conformal derivative $\cp_{\m}$, formed from the trace-free Christoffel symbols $\widetilde{\ch}$ associated with the conformal metric $\cg_{\m\n}$, satisfying

\be
\label{eq.cgc}
\cp_{\m}\cg_{\kappa\la}=0;
\ee

and the metric derivative $\metp_{\m}$, formed from the Christoffel symbols $\ch$ associated with $\g_{\m\n}=e^{\varphi}\cg_{\m\n}$, satisfying
\be
\metp_{\m}\g_{\ka\la}=0.
\ee

The relation between the two is
\be
\label{eq.mcequiv}
\metp_{\m}\g_{\ka\la}=e^{\varphi}\cp_{\m}\cg_{\ka\la}.
\ee

From these two connections, we can construct two curvature tensors: the metric curvature $\tens{K}^{\ddd\ka}_{\n\m\la}$ and conformal-connection curvature $\widetilde{\tens{C}}^{\ddd\ka}_{\n\m\la}$,  formed using $\ch$ and $\widetilde{\ch}$ respectively. $\widetilde{\tens{C}}^{\ddd\ka}_{\n\m\la}$ is not to be confused with the usual Weyl curvature tensor $\tens{C}^{\ddd\ka}_{\n\m\la}$ which can be written as a function of $\widetilde{\tens{C}}^{\ddd\ka}_{\n\m\la}$ and $\widetilde{\tens{C}}_{\m\la}$.

\item{Projective structure and affine one-form}

In standard GR, the affine connection $\G_{\m\n}^{\s}$ can be split into a trace-free geometric object with components $\tPi_{\m\n}^{\s}$  and a trace $\G_{\m}$. Equivalence classes of projectively equivalent connections define the projective structure corresponding to $\tPi_{\m\n}^{\s}$. This structure, together with a direction at any point of $\M$, determines an autoparallel path, i.e. a path whose tangent direction is transported into itself under parallel transport by any member of the equivalence class. Physically, these correspond to the paths of massive, acceleration-free, structureless particles~\cite{Weyl1921}. Classically, the particle paths are primary; but, by considering an ensemble of such paths defined by a complete solution to the relativistic Hamilton-Jacobi equation, we can go over to a wave-front description of the ensemble~\cite{Benenti}. 

The equivalence class of projectively equivalent connections can be generated by forming the ``symmetrized sum" of $\tPi^{\s}_{\m\n}$ and an arbitrary one-form $\tens{\Theta}_{\m}$\cite{Schouten1954},
\be
\label{eq.b}
\G^{\ka}_{\m\n}=\tPi^{\ka}_{\m\n}+\f{1}{5}\left(\tdel^{\ka}_{\m}\tens{\Theta}_{\n}+\tdel^{\ka}_{\n}\tens{\Theta}_{\m}\right).
\ee
By choice of different one-forms, one generates the equivalence class of all connections projectively equivalent to $\tPi^{\s}_{\m\n}$. A particular choice of a one-form field $\tens{\Theta}_{\m}$ determines the prefered affine parameter along paths of free massive particles and a unique affine connection. Under $\GL$, the transformation laws of $\tPi_{\m\n}^{\s}$ and $\G_{\m}$ are not trivial. Under $\SL$, each equivalence class $\tPi^{\ka}_{\m\n}$ corresponds to a trace-free projective connection, while $\G_{\m}$ is a one-form and so $\tens{\Theta}_{\m}=\G_{\m}=\G^{\ka}_{\m\ka}$.

There are two covariant derivatives associated with the projective structure: the projective derivative $\pp_{\m}$ formed using $\tPi^{\s}_{\m\n}$, and the affine derivative  $\ap_{\m}$ formed using $\G^{\ka}_{\m\n}$.  Their respective curvatures tensors, the projective-connection curvature $\tPi^{\ddd\ka}_{\n\m\la}$ and affine curvature $\Ri^{\ddd\ka}_{\n\m\la}$ are formed using $\tPi^{\ka}_{\m\n}$ and $\G^{\ka}_{\m\n}$ respectively. 
$\tPi^{\ddd\ka}_{\n\m\la}$ is not to be confused with the usual projective curvature tensor $\tens{P}^{\ddd\ka}_{\n\m\la}$, which can be written as a function of $\tPi^{\ddd\ka}_{\n\m\la}$ and $\tPi_{\m\la}$. We discuss some implications of the distinction between the projective and conformal curvature tensors for quantum gravity theory in the concluding section.

The affine curvature tensor gives rise to the Ricci tensor $\Ri_{\m\n}$ which can be split into a symmetric part $\Ri_{(\m\n)}$ and a two-form $\Ri_{[\m\n]}$. If $\Ri_{\m\n}$ is symmetric, the connection is said to be volume-preserving. Then, locally, $\G = \tens{d}\tens{\Psi}$, where $\tens{\Psi}$ is a scalar field. 
\end{itemize}

\section{Compatibility conditions in UCPR}

In UCPR, full compatibility between the metric and affine structures may be approached in a series of steps. Their compatibility may be ``measured" by the tensor $\td^{\dd\ka}_{\m\n}$ which is equal to the difference between the connection $\G^{\ka}_{\m\n}$ and the Christoffel symbols $\ch$ corresponding to the metric, and can be written as~\cite{Schouten1954}
\be
\label{eq:tdef}
\td^{\dd\ka}_{\m\n}=\f{1}{2}\g^{\ka\s}\left(\q_{\m\s\n}-\q_{\s\n\m}+\q_{\n\m\s}\right)
\ee
where 
\be
\q_{\m\s\n}\equiv-\nabla_{\m}\g_{\s\n}.
\ee

In UCPR, we also define the projective-conformal analogue of  $\td^{\dd\ka}_{\m\n}$, which is equal to the difference between the projective and conformal connections, $\tPi^{\ka}_{\m\n}-\widetilde{\ch}$, and can be written as
\be
\label{eq:tdef}
\cpd^{\dd\ka}_{\m\n}=\f{1}{2}\cg^{\ka\s}(\cpq_{\m\s\n}-\cpq_{\s\n\m}+\cpq_{\n\m\s}),
\ee
where 
\be
\cpq_{\m\s\n}\equiv-\pp_{\m}\cg_{\s\n}.
\ee
The relation between the two $\tens{T}$'s is 
\be
\label{eq.tcpt}
\td^{\dd\ka}_{\m\n}=\cpd^{\dd\ka}_{\m\n}+\frac{2}{5}\delta^{\ka}_{(\m}\G_{\n)}-\delta^{\ka}_{(\m}\p_{\n)}\varphi+\f{1}{2}\cg_{\m\n}\cg^{\ka\s}\p_{\s}\varphi.
\ee
Full compatibility between the metric and affine structures is assured by the vanishing of $\td^{\dd\ka}_{\m\n}$.
UCPR allows us to define various partial compatibility conditions by imposing certain restrictions on the $\tens{T}$'s. We give a few obvious examples.
\subsection{Equi-affine condition}
As is well known, the four-volume element $\ep$ is invariant under parallel transport by the affine connection if $\G_{\m}=2\p_{\m}\varphi$~\cite{Veblen, Sanchez2003}. This  \emph{equi-affine condition} can be expressed as
\be
\label{eq:equiaffine}
\td^{\dd\ka}_{\m\ka}=\G_{\m}-2\p_{\m}\varphi=0,
\ee
or
\be
\td^{\dd\ka}_{\m\n}=\cpd^{\dd\ka}_{\m\n}-\frac{1}{5}\delta^{\ka}_{(\m}\p_{\n)}\varphi+\f{1}{2}\cg_{\m\n}\cg^{\ka\s}\p_{\s}\varphi.
\ee
Given any projective connection, the equi-affine condition fixes a unique affine connection~\cite{Mikes}. However, the equi-affine condition does not limit the relation between the confomal and projective structures. 
\subsection{Weyl Compatibility Condition}
A Weyl space is one in which
\be
\label{eq.twcond}
\td^{\dd\ka}_{\m\n}=\f{1}{2}\left(\q_{\m}\delta^{\ka}_{\n}+\q_{\n}\delta^{\ka}_{\m}-\cg_{\m\n}\cg^{\ka\rho}\q_{\rho}\right),
\ee
where $\q_{\m}$ is an arbitrary one form~\cite{Schouten1954}. Taking the trace of this expression over $\ka$ and $\n$, we see that
\be
\q_{\m}=\G_{\m}/2-\p_{\m}\varphi.
\ee
Since the Weyl condition imposes no restrictions on the relation between $\G_{\m}$ and $\varphi$, we can express it entirely in terms of $\cpd^{\dd\ka}_{\m\n}$:
\be
\label{eq.wcpt}
\cpd^{\dd\ka}_{\m\n}=\f{1}{4}\left(\frac{1}{5}\delta^{\ka}_{\m}\G_{\n}+\frac{1}{5}\delta^{\ka}_{\n}\G_{\m}-\cg_{\m\n}\cg^{\ka\s}\G_{\s}\right).
\ee

The Weyl condition ensures that a null vector field $\tens{k}^{\m}$, with $\tens{k}_{\m} = \p_{\m}\Omega$, obeying the null Hamilton-Jacobi equation~\cite{Benenti}
\be
\label{eq.nhj}
 \cg^{\m\n}\tens{k}_{\m}\tens{k}_{\n}=0,
\ee
also obeys the equation~\cite{Ehlers1972}:
 \be
\tens{k}^{\m}\pp_{\m}\tens{k}^{\ka}=\f{1}{5}\G_{\m}\tens{k}^{\m}\tens{k}^{\ka}.
\ee
In other words, the Weyl condition ensures that conformal null geodesics are also projective autoparallel paths. 

\subsection{Full conformal-projective compatibility}

The condition for full conformal-projective compatibility, which does not seem to have been considered so far, is
\be
\label{eq.cpeq}
\cpd^{\dd\ka}_{\m\n}=0.
\ee
If the conformal and projective structures are compatible,  \emph{all} vector fields $\tens{v}^{\m}$, with $\tens{v}_{\m} = \p_{\m}\psi$, obeying Hamilton-Jacobi equation~\cite{Benenti}
\be
\label{eq.thj}
\cg^{\m\n}\tens{v}_{\m}\tens{v}_{\n}= 0, \pm1,
\ee
whether null, timelike or spacelike, are also tangent to projectively autoparallel paths; i.e. they obey
\be
\tens{v}^{\m}\pp_{\m}\tens{v}^{\ka}\propto \tens{v}^{\ka}.
\ee
So, if the full compatibility between the conformal metric and the projective connection holds, then it follows that all conformal geodesics are also autoparallel paths.

\subsection{Full metric-affine compatibility}

The usual expression for the full metric-affine compatibility is that the affine derivative of the metric tensor vanish, which is to say that the affine connection be identical to the metric connection, or 

\be
\label{iii}
\td^{\dd\ka}_{\m\n}=0.
\ee
This implies not only that all conformal geodesics also be autoparallel paths, but that the preferred parameter of a geodesic also be a preferred parameter of the corresponding autoparallel curve. Full metric affine compatibility is ensured by demanding that both the Weyl condition \emph{and} the equi-affine condition hold. 

\section{Lagrangian formulation of UCPR}
By treating $\varphi$, $\cg^{\m\n}$, $\G_{\m}$, and $\proj$ as four independent fields in a Palatini-type variational principle, one can derive the vacuum and non-vacuum field equations and the compatibility conditions.

The usual GR Lagrangian can be written as a function of $\g^{\m\n}$ and $\G^{\s}_{\m\n}$
\be
\label{eq.lagrangian}
\mathcal{L}=\sqrt{-\g}\g^{\m\n}\Ri_{\m\n}+\ka\T^{\m\n}\g_{\m\n},
\ee
where $\T^{\m\n}$ is the stress-energy tensor density, here assumed to be independent of the metric tensor~\cite{Stachel1969}. Its variation results in ten field equations for the connection and forty compatibility conditions between the metric and the connection. We can rewrite this Lagrangian as 
\be
\mathcal{L}=e^{\varphi}\left(\cg^{\m\n}\Ri_{\m\n}+\ka\T^{\m\n}\cg_{\m\n}\right),
\label{eq:l}
\ee
where $\tens{R}_{\m\n}$ is equal to 
\be
\tens{R}_{\m\n}=\tPi_{\m\n}+\f{2}{5}\p_{[\n}\G_{\m]}-\f{3}{5}\left(\pp_{\m}\G_{\n}-\f{\G_{\m}\G_{\n}}{5}\right).
\label{eq:rpiriccit}
\ee

After canceling the common $\ep$ factors, the field equations resulting from variation of all four fields are
\begin{subequations}
\be
\label{eq:phife}
\widetilde{\Ri}=-\ka\widetilde{\T};
\ee
\be
\label{eq:cgfe}
\Ri_{\m\n}-\f{1}{4}\cg_{\m\n}\widetilde{\Ri}=\ka(\T^{\s\rho}\cg_{\s\m}\cg_{\rho\n}-\f{1}{4}\cg_{\m\n}\widetilde{\T});
\ee
\be
\label{eq:Gfe}
\cg^{\m\n}\td^{\ka}_{\m\n}=0\tx{; and }
\ee
\be
\label{eq:Pife}
\cg^{\m\n}\td^{\rho}_{\s\rho}-2\td_{\s\rho}^{\dd(\m}\cg^{\n)\rho}+\f{2}{5}\cg^{\ka\rho}\td_{\ka\rho}^{\dd(\m}\delta^{\n)}_{\s}=0.
\ee
\end{subequations}
Here $\widetilde{\Ri}=\cg^{\m\n}\Ri_{\m\n}$ and $\widetilde{\T}=\T^{\m\n}\cg_{\m\n}$.

Equation~(\ref{eq:cgfe}) is identical to the field equation of unimodular relativity;  eq. ~(\ref{eq:cgfe}) and eq. (\ref{eq:phife}) together are equivalent to the field equations of GR. Combining  (\ref{eq:Gfe}) and (\ref{eq:Pife}) yields
\be
\label{eq:PiGfe}
\cg^{\m\n}\td^{\rho}_{\s\rho}-\td_{\s\rho}^{\dd\m}\cg^{\n\rho}-\td_{\s\rho}^{\dd\n}\cg^{\m\rho}=0.
\ee
The $\cg_{\m\n}$ trace of this expression is equivalent to the equi-affine condition in eq.~(\ref{eq:equiaffine}). The $\cg_{\m\n}$-trace-free part
\be
\label{eq:PiGfeTraceFree}
\f{1}{2}\td^{\rho}_{\s\rho}-\td_{\s\rho}^{\dd\m}\cg^{\n\rho}-\td_{\s\rho}^{\dd\n}\cg^{\m\rho}=0
\ee
is satisfied if the Weyl condition in  eq.~(\ref{eq.twcond}) holds. Together, they are equivalent to the full compatibility condition.

More interestingly, the treatment of the four fields Êas independent enables us to truncate the standard GR Lagrangian in various ways. For example,  setting $\varphi=const.$ and $\G_{\m}=0$, we get a conformal-projective variational principle, the result of which are two field equations:
\begin{subequations}
\be
\ep\left(\tPi_{\m\n}-\f{1}{4}\cg_{\m\n}\widetilde{\tPi}\right)= \ka\ep\left(\T^{\s\rho}\cg_{\s\m}\cg_{\rho\n}-\f{1}{4}\cg_{\m\n}\widetilde{\T}\right)
\ee
\mbox{and}
\be
\cpd^{\dd\s}_{\m\n}=0.
\ee
\end{subequations}
This case may be an appropriate description of the space-time geometry in theories in which only massless fields are present. 

In the other extreme case, the projective and conformal structures are fixed and compatible, i.e. $\cpd^{\dd\ka}_{\m\n}=0$, while $\varphi$ and $\G_{\m}$ are dynamical. The variation of such a Lagrangian yields
\begin{subequations}
\be
\widetilde{\tens{R}}=-\ka\widetilde{\T}
\ee
\mbox{and}
\be
\cg^{\m\n}\left(\p_{\m}\varphi+\f{2}{5}\G_{\m}\right)=0.
\ee
\end{subequations}

And of course, there is no reason why one must restrict attention to these two cases. Many other Lagrangians invariant under $\SL$ can be formed from the four fields discussed, and some of these have already been considered in alternative gravitational theories~\cite{Fatibene}. In~\cite{Bradonjic} we shall discuss this in greater detail and also consider the much-discussed question of the inclusion of the cosmological constant in the Lagrangian.

\section{Conclusion}
Our discussion of conformal and projective structures raises some questions for further investigation at the quantum level~\cite{Stachel2011}. First of all, one may wonder about the central role played by $\SL$. Compared to $\GL$, the new element it introduces is an invariant four-volume structure at each point of space-time. It has been suggested that this structure may be the remnant, in the classical continuum limit, of a more fundamental discrete quantum structure of space-time itself~\cite{Anderson}. 
Rather than simply being postulated, as in causal set theory~\cite{Daughton}, this discreteness itself may be the result of some dynamical quantization procedure.  Here, we restrict our further discussion to the role of a four-volume element in the quantization of other space-time structures, based on the assumption of an underlying differentiable manifold.

An analysis of the electro-magnetic field provides an instructive example. Bohr and Rosenfeld emphasized complementarity between the conditions required for the definition and measurement of the field strengths and phases, and the conditions required for counts of photons with given energies and momenta~\cite{Bohr, Stachel2009}. Bergmann and Smith have discussed an important distinction between the analysis of the field in space-time and its analysis in energy-momentum-space: in the former, a four-dimensional region of integration is needed to define average values, while in the latter, a three-dimensional region of a null hypersurface is needed~\cite{Bergmann}. 

The average value of all components of the electromagnetic field tensor in any region of space-time  can be defined and measured with arbitrary accuracy.  Bergmann and Smith, and DeWitt have discussed the measurement of the physical components of the gravitational curvature tensor by methods analogous to those of Bohr and Rosenfeld~\cite{Bergmann, DeWitt}. They work in empty space-times, so it is not clear which form of the curvature tensor is involved; but we suspect that it is the projective curvature tensor associated with the fields generated by non-zero rest mass sources. Ehlers and Schild, and Pirani have shown how this curvature tensor can be defined and measured by projectively invariant techniques~\cite{Ehlers1973,Pirani1971}. Again, one should be able to treat interacting electromagnetic and gravitational fields by some generalization of the approaches of Bergmann and Smith, and DeWitt. UCPR provides a framework suitable for such generalization. Indeed, one can investigate the field at timelike projective infinity~\cite{Eardley1973,Schmidt}. There are also techniques for uniting the approaches to null, timelike and spatial infinity~\cite{Soleimany}. 

Theoretically, the``free" radiation field that has escaped from its sources must be treated in terms of light quanta (photons) to account for the discrete events registered by devices sensitive to individual field quanta. Of course, one can formally Fourier analyze any electromagnetic field and interpret the result in terms of field quanta; but that does not imply that the resulting Fock-space occupation numbers will necessarily be observables under all conditions of measurement. 

The Bondi-Metzner-Sachs field of any bounded radiating source~\cite{Sachs1962} asymptotically approaches the ``free" gravitational radiation field, which is represented by the conformal structure at null infinity as defined by Penrose~\cite{Penrose}. This field has been quantized by techniques anticipated by Komar~\cite{Komar} and developed in full detail by Ashtekar~\cite{Ashtekar}. The resulting asymptotic gravitons are representations of the Bondi-Metzner-Sachs group.

One can also treat ``pure" radiation fields independently of their sources; again one needs only conformal structures to formulate the results mathematically, interpret them physically, and describe procedures for their measurement~\cite{Pirani1961, Pirani1961b, Pirani1971, Ehlers1972, Kristian, Ellis1969, Ellis1985}. As discussed in detail by Reisenberger, it should thus be possible to quantize these fields directly~\cite{Reisenberger2007, Reisenberger2008}. The homogeneous Maxwell equations are conformally invariant, so one should also be able to treat interacting ``free" electromagnetic and gravitational fields by conformal techniques.

Much of recent work on spin networks in loop quantum gravity is based on the holonomy group of the affine connection for loops on a space-like hypersurface. Currently, the four-dimensional spin foam amplitude connects spin networks on initial and final spacelike hypersurfaces ~\cite{Alexandrov, Oriti}. But in order to discuss the scattering of zero rest mass fields, such as the electromagnetic or gravitational, spin foam amplitudes should connect spin networks on initial and final null hypersurfaces. In particular, for asymptotically flat fields, it should connect those on future and past null infinity~\cite{Penrose}. Consideration of the holonomy group of the connection that satisfies the full conformal-projective compatibility condition may open up new possibilities for the construction of four-dimensional null spin foam amplitudes. 

\acknowledgments
J. S. gratefully acknowledges earlier work on conformal and projective structures with Dr. Mihaela Iftime.

\end{document}